# How do biofilms feel their environment?

Merrill Asp[1,2], Minh Tri Ho Thanh[1,2], Arvind Gopinath[3,4] and Alison Patteson[1,2]

1 Physics Department, Syracuse University

2 BioInspired Institute, Syracuse University

3 Department of Bioengineering, University of California, Merced

4 Health Sciences Research Institute, University of California, Merced

**Abstract:** The ability of bacteria to colonize and grow on different surfaces is an essential process for biofilm development and depends on complex biomechanical interactions between the biofilm and the underlying substrate. Changes in the physical properties of the underlying substrate are known to alter biofilm expansion, but the mechanisms by which biofilms sense and respond to physical features of their environment are still poorly understood. Here, we report the use of synthetic polyacrylamide hydrogels with tunable stiffness and controllable pore size to assess physical effects of the substrate on biofilm development. Using time lapse microscopy to track the growth of expanding *Serratia marcescens* colonies, we find that biofilm colony growth can increase with increasing substrate stiffness on purely elastic substrates, unlike what is found on traditional agar substrates. Using traction force microscopy, we find that biofilms exert transient stresses correlated over length scales much larger than a single bacterium. Our results are consistent with a model of biofilm development in which the interplay between osmotic pressure arising from the biofilm and the poroelastic response of the underlying substrate controls biofilm growth and morphology.

**Introduction**

The growth and expansion of biofilms over new surface territories is a fundamental step in the biofilm development process. This process depends on complex signaling pathways that link the cell to its extracellular environment. When a bacteria makes contact with a surface, it initiates a program of gene expression that promotes colonization and biofilm formation (*1, 2*). As the biofilm grows, it absorbs nutrients from its environment and generates internal forces that allow the biofilm to expand and move against its substrate (*3-6*). Biofilm growth thus depends on chemical signals but also mechanical interactions with its environment. While there is strong evidence that bacteria can sense and respond to physical features of their environment, the mechanisms of surface sensing remain poorly defined.

Bacteria colonize a staggering range of complex environment from soil and waterways to biological fluids and tissues. An emerging number of studies reveal multiple and often interconnected ways in which physical signals from the environment affect biofilm development. For example, recent investigations show that biofilm expansion is enabled by the production of extracellular polymers that induce osmotic pressure gradients and drive the flow of fluid and nutrients from a hydrogel substrate into the biofilm (*7-10*). The substrate pore size and its effects on fluid permeability acts as a limiting factor to biofilm growth. On the other hand, a number of studies indicate an important role of substrate stiffness on biofilm by mediating adhesion and frictional forces between the biofilm and the substrate (*3, 11-13*). Together, these works highlight a critical interplay between biofilm growth and the material properties of the substrate and the need to decipher multiple physical signals to understand biofilm growth in complex environments.

Here we report the development of polyacrylamide (PAA) hydrogels with tunable stiffness and matrix porosity to determine their integrated effects on biofilm growth. Polyacrylamide gels have been widely used in mammalian culture for the last couple decades to study how cells sense and response to their local environment (*14-16*) but are only beginning to be used for bacteria culture (*11, 17*). PAA gels offer chemical and mechanical advantages compared to the most common bacteria growth substrate, agar. PAA forms well characterized linearly elastic gels, which isolates the effects of substrate stiffness on biofilm growth from the non-linear rheology and viscoelastic energy dissipation present in agar (*18, 19*). This allows facile calculations of stresses from measurements of the substrate displacements, which provide directly the forces exerted by the biofilm on the substrate. Furthermore, unlike agar and most other bacteria growth substrates, the shear moduli of PAA gels can be tuned by either crosslinker concentration or polymer concentration, which allows independent tunability of matrix stiffness and matrix pore size (*20*). This feature allows modification of the elastic moduli of the substrate while keeping pore size nearly constant, thus disentangling effects of stiffness and nutrient transport.

In this manuscript, we show that in the limit of purely elastic substrates *Serratia marcescens* colonies spread out faster on stiffer substrates compared to softer ones. This is the opposite of their behavior on soft agar substrates. Using traction force microscopy, we find that biofilms exert transient stresses correlated over length scales much larger than a single bacterium. Our results are consistent with a model of biofilm development that arise from osmotic pressure gradients between the biofilm and the substrate and the substrates poroelastic response. Our results demonstrate the use of polyacrylamide hydrogels as a model platform for quantifying biofilm growth and their response to physical changes in their environment.

**Results**

***Design and characterization of polyacrylamide hydrogels***

In this study, we used both polyacrylamide gels and conventional agar as a point of comparison. To characterize the mechanical properties of the gels, we measured their elastic modulus $G'$, which quantifies their resistance to shear deformations, and a loss modulus $G''$, which quantifies viscous energy dissipation, with an oscillatory rheometer. As shown in Fig. 1a, agar exhibits non-linear shear softening, decreasing from approximately 10 to 1 kPa as shear strain rises from 2 to 50%. This suggests that biofilms might sense different agar stiffness depending on the extent to which the biofilms deform the agar substrate. Agar is also viscoelastic, with a viscous loss modulus $G''$ of 10%-50% of the storage modulus $G'$ at least for small strain values (2-5%). This data suggests that as a substrate for biofilm growth agar dissipates energy and relaxes applied stresses that might be relevant to outward growth of the colony. The mechanical response of agar to shear strain differs from PAA gels. As shown in Fig. 1b, polyacrylamide gels form linearly elastic gels. PAA gels also exhibit negligible viscous dissipation, consistent with previous work (*21-23*).

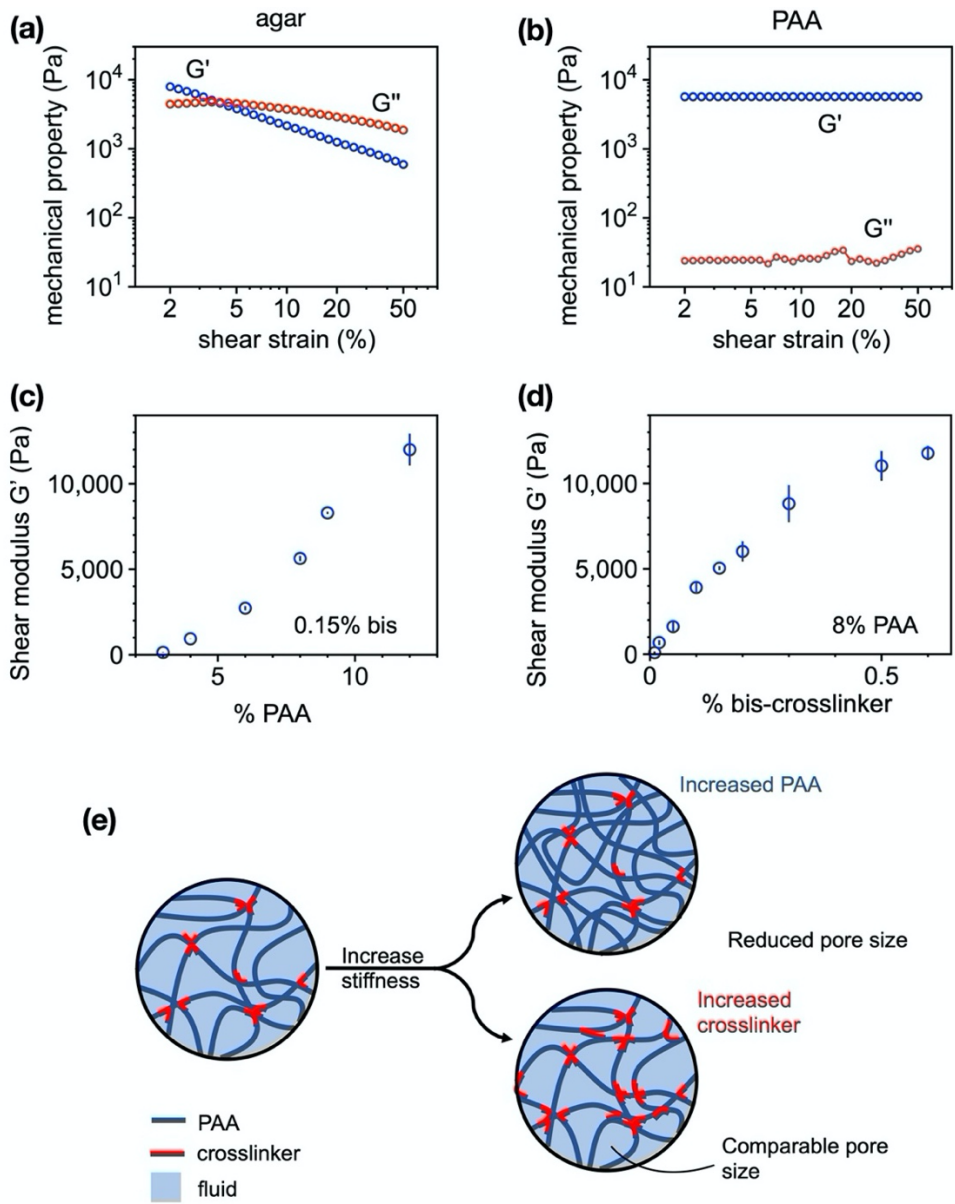

**Fig.1.** Substrate characterization using a stress-controlled rheometer. (a,b) The shear storage modulus G' and loss modulus G'' as a function of shear strain for both agar and polyacrylamide gels. (c,d) The shear storage modulus of PAA gels are prepared by increasing PAA concentration and chemical crosslinker bis-acrylamide. (e) Schematic representation illustrating the effects of increasing PAA polymer concentration vs chemical crosslinker on fluid permeability in the network.

In order to distinguish between substrate stiffness and substrate permeability effects on biofilm growth, we designed PAA gels with shear moduli G' ranging from 100-10,000 Pa by varying either the amount of acrylamide or the amount of chemical crosslinker bis (Fig. 1c&d). These two parameters, acrylamide concentration and crosslinker concenter, have two different effects on network permeability (Fig. 1e). Increasing the concentration of acrylamide monomer results in a denser, stiffer polyacrylamide network with a smaller pore size and lower permeability. Increasing the concentration of crosslinker links together the same density of polyacrylamide polymers at a greater number of sites, increasing the network stiffness without significantly changing pore size. These effects are illustrated schematically in Fig. 1e. We independently verified the expected changes in network permeability by directly tracking the diffusion of dyed fluid into PAA gels of varying compositions of acrylamide and bis crosslinker (SI). While substrates prepared by increasing polyacrylamide concentration reduced the spread and diffusion of the dye into the cell, for substrates of varying bis-crosslinker, the dyes diffused at the same rate, suggesting that increased bis enhances network stiffness yet holds the network pore size and fluid permeability relatively constant.

### *Substrate stiffness increases biofilm expansion rates*

Our experimental protocol consists of directly observing the growth of *Serratia marcescens* colonies on the surface of hydrogel substrates with time-lapse microscopy (Methods). Before inoculation, the PAA gels are soaked multiple times in LB nutrient broth, the same fluid component of agar gels used here. We deposit a small inoculum of bacteria on the gel surfaces and track x-y positions of the resulting biofilm boundary as it expands over time $t$ (Fig. 2).

Representative biofilm time lapse images on a soft (G' = 0.9 kPa) and a stiff (G' = 3 kPa) PAA gel are shown in Fig. 3. Strikingly, biofilm expansion is faster on the stiffer PAA gel compared to the softer one. This behavior is opposite to that on conventional agar substrates, where colony growth is slower on stiffer agar compared to softer agar. This phenomenon is highlighted in Figure 4a, which shows snapshots of whole biofilm colonies on PAA gels and agar substrates taken at 15 hr post inoculation. We note that the reduced biofilm growth on stiff agar substrates compared to soft agar is a common feature of many different types of bacteria species (*7, 24-26*). This inhibited biofilm growth on agar has been attributed to the reduction in substrate permeability found in more concentration agar substrates, which limits the transport of fluid and nutrients from the substrate into the biofilm. The strong increase in biofilm expansion in more concentrated PAA gels is thus unexpected (Fig. 3&4).

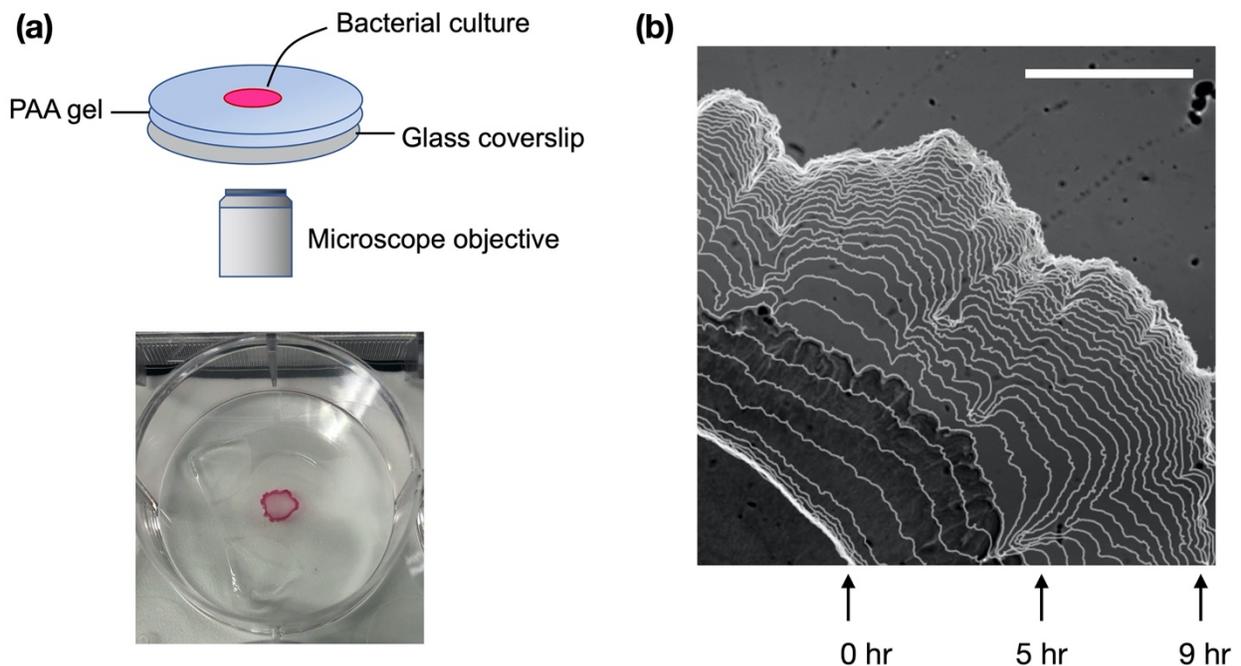

**Fig.2. (a)** Schematic diagram of the experimental setup**.** Aliquots of *Serratia marcescens* were placed on polyacrylamide (PAA) substrates that were 1 mm in height. The substates were maintained at 37 °C in a humid stage top incubator. **(b)** Visualization of the growing biofilm boundary overlaid on a sample image of the biofilm. Images were acquired at 10 minute increments. Scale bar, 1 mm.

To quantify the above observations, we calculate the initial biofilm expansion rates by calculating the boundary velocities from the tracking data. The biofilm velocity is defined here as the average radial displacement of the biofilm boundary over a time interval of $\Delta t$ = 20 min. Figure 4b&c show initial biofilm velocities on PAA and agar substrates taken at 2 hr post inoculation, when radial expansion of the biofilms are first beginning to be observed. The data shows that for small G' values (G' < 5 kPa) biofilm velocity increases with substrate stiffness on our purely elastic substrates as prepared by varying acrylamide or varying bis crosslinker. For G'> 5 kPa, biofilm velocity seems to saturate with substrate stiffness and perhaps begins to slowly decrease on PAA gels. These results are entirely different from biofilm growth on agar substrates, which decreases dramatically with increased agar concentration (Fig. 4c). Our results indicate that on purely elastic substrates, substrate stiffness increases biofilm expansion rates. This is most evident in the varying bis data in which the network permeability is approximately constant (SI). Differences between varying PAA and varying bis preparations could point to effects of network permeability, which differs between the two preparations. Interestingly, the varying PAA and varying bis data differs most on soft substrates (G' < 5 kPa), in which biofilms are most sensitive to substrate stiffness.

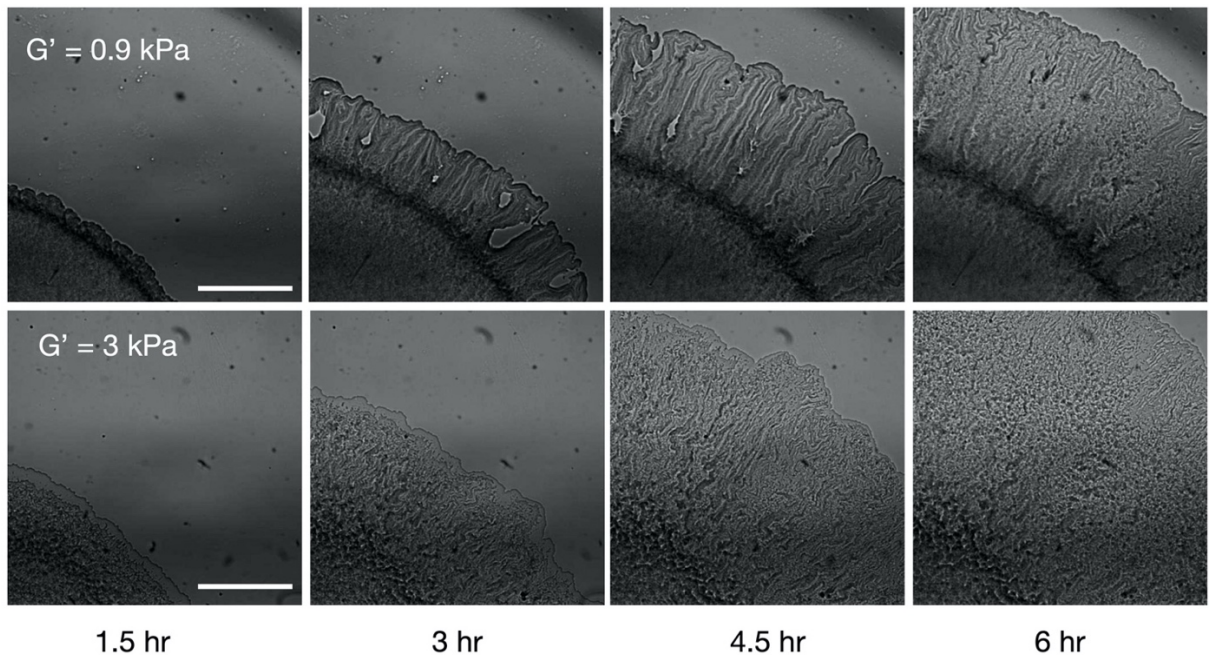

**Fig.3.** Representative time-lapse images of *Serratia marcescens* colonies growing across a soft (0.9 kPa) and stiff (3 kPa) PAA gel. Scale bar, 1 mm.

Taking into account the diverse bacterial strains that colonize agar, we selected three additional bacterial cell lines to test on PAA gels. These cell lines were *Pseudomonas aeruginosa*, *Proteus*

*mirabilis*, and *Myxococcus xanthas*. Figure 3 shows biofilm expansion rates on a soft (G' = 0.5 kPa) and stiff (G' = 5 kPa) PAA substrate. In all the cases we investigated with these different cell lines, biofilm expansion was faster on stiffer PAA than softer PAA.

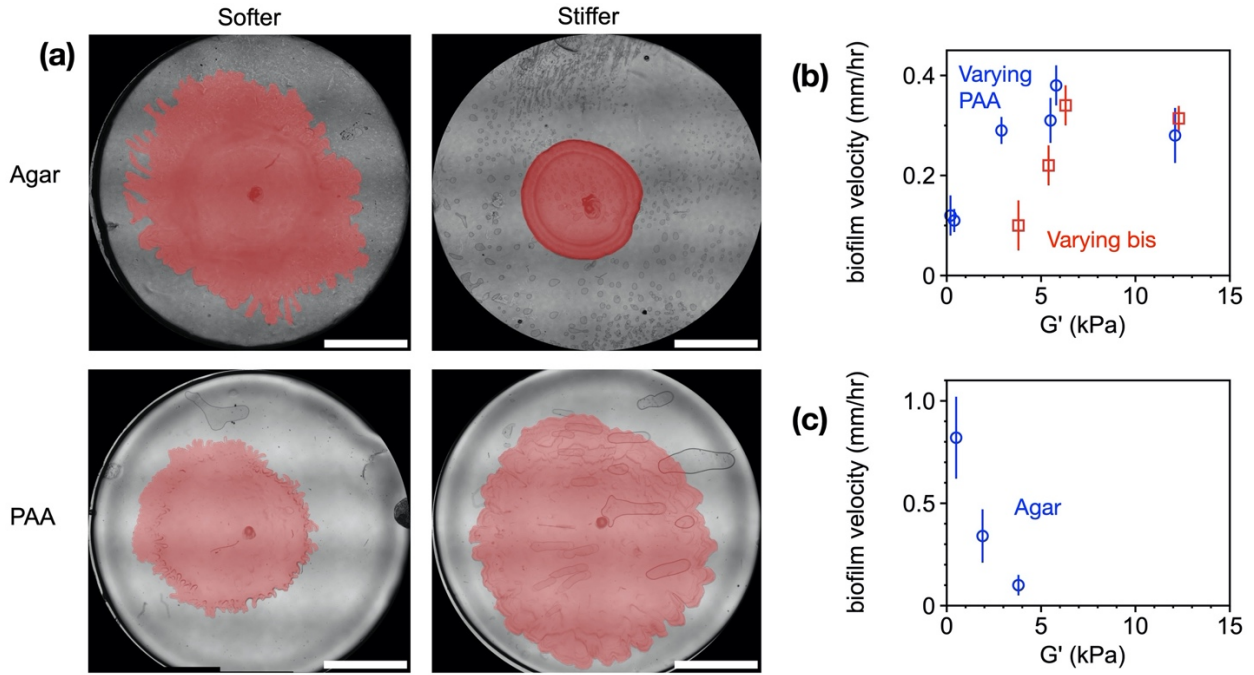

**Fig.4. (a)** Representative whole-colony pictures of *Serratia marcescens* grown on soft and stiff agar and PAA substrates. While colonies size decreases on stiffer agar substrates, the opposite occurs on purely elastic PAA substrates. The biofilms are manually traced and pseudo-colored pink to enhance imaging contrast. Scale bar, 5 mm. **(b&c)** Biofilm expansion velocity as a function of substrate stiffness (G') for *Serratia marcescens* colonies grown on (b) PAA and (c) agar substrates. For agar, G' is chosen at the shear stiffness in the limit of 0% shear strain.

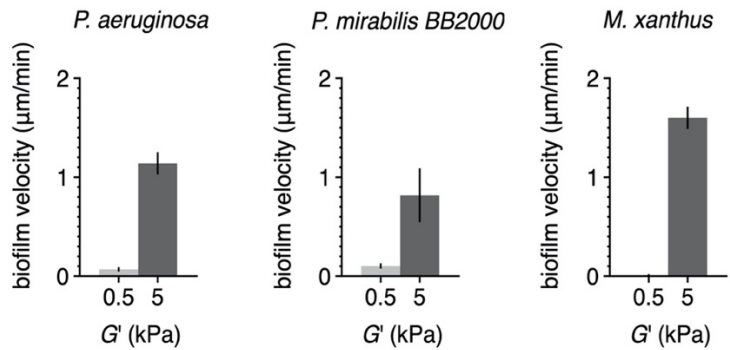

**Fig.5.** Biofilm expansion rates on soft (G' = 0.5 kPA) and stiff (G' = 5 kPa) PAA gels for *P. aeruginosa*, *P. mirabilius*, and *M. xanthus* bacteria strains.

*Biofilm force generation and associated substrate deformations*

What may cause substrate elasticity to increase biofilm expansion rates? To better understand the observed enhancement in biofilm expansion with substrate stiffness G', we performed experiments in which biofilm-generated substrate displacements could be directly observed via traction force microscopy techniques (Fig. 6). In these experiments, the polyacrylamide gels were embedded with fluorescent marker beads with an average bead-to-bead spacing of 5 µm. The displacements of these beads were tracked in the vicinity of the expanding edge of the biofilm as it moved outward over time. On soft 850 Pa gels, we observed a pulse of substrate deformations associated with the outward movement of the *Serratia marcescens* biofilm edge (SI Video). Strikingly, some of the substrate deformations occurred out in front of the biofilm edge, pushing out in regions in which the biofilm had not yet even spread to (Fig. 6a). Based on the stiffness of substrate, the estimated traction stresses exerted by the bacteria on the hydrogel surface were on the order of 50-100 Pa.

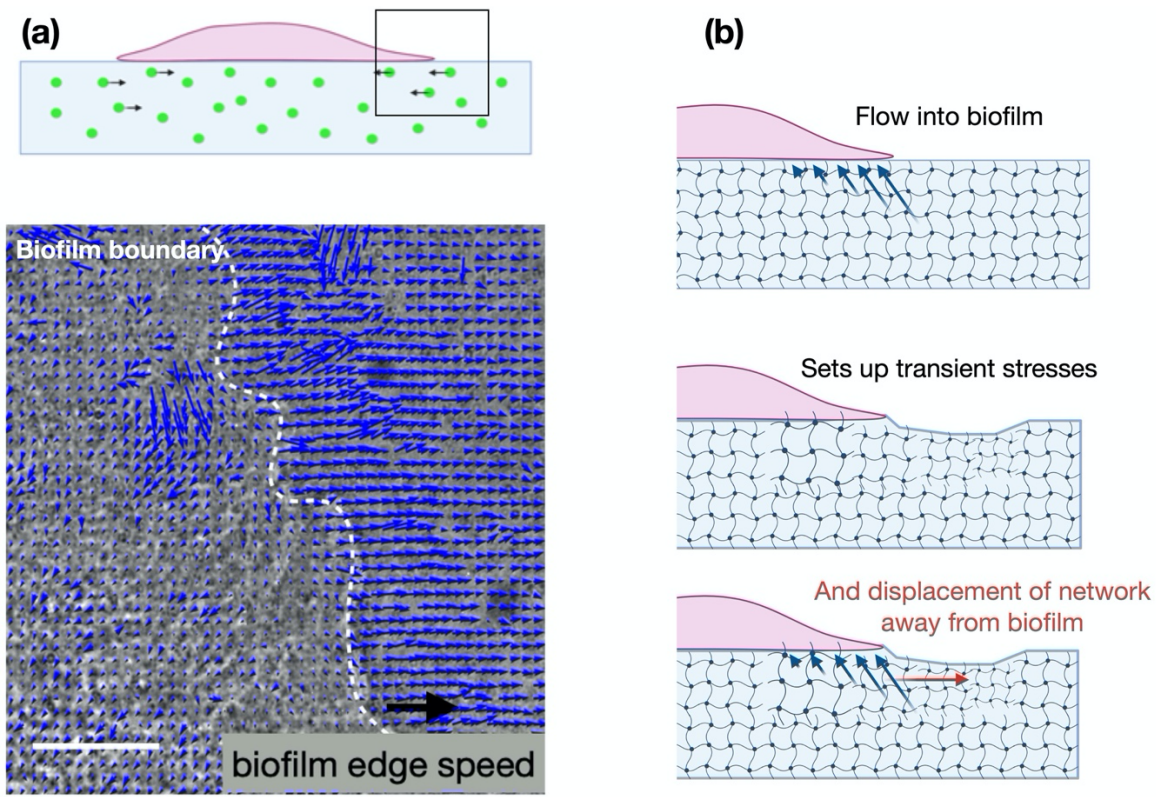

**Fig. 6. Traction force microscopy experiments. (a)** Here, we embedded a 850 Pa PAA gel with fluorescent beads to trace substrate deformations in the vicinity of a growing biofilm front. We observe a pulse of substrate displacements that travel out with the expanding edge (SI Video). Notable, substrate deformations occur out in front of the leading biofilm edge. Hotspots of network displacements correspond to 50-100 Pa stresses. Scale bar, 50 µm. **(b)** Schematic representation of poroelastic stresses associated with a growing biofilm front.

To interpret these results (Fig. 6), we suggest a minimal model that treats substrate deformations as a signature of poroelastic stresses in the network driven by osmotic pressure differences from the growing biofilm front. In this picture, the hydrogel substrate behaves as a poroelastic material, compromising of a porous elastic network component permeated by a fluid that can move relative to this network. Biofilm growth commences as bacteria begin to excrete extracellular polymers. Before these polymers assemble into extracellular matrix network, they act as osmolytes that set up an osmotic pressure difference between the biofilm and the substrate (*7-10*). Gradients in osmotic pressure draw up fluid and nutrients into the biofilm, which allows the biofilm front to grow and expand.

Here we propose that this fluid flow sets up transient stresses in the substrate network, which could drive substrate displacements in regions surrounding the biofilm (schematic in Fig. 6b). In a poroelastic material, fluid flows and network deformations are coupled. Based on our experimental observations (Fig. 4&6), we infer that stiffer substrate networks more efficiently couple with fluid flows, increasing transmission of forces through the network and driving faster expulsion of the fluid out of the network. On soft substrates, in contrast, local strains decay faster, resulting in reduced propagation and transmission of stress. The flow of fluid to relax the applied stress is thus localized to smaller regions resulting in reduced fluid and nutrient flux. In this way, substrate network stiffness may act to increase initial biofilm growth rates.

Taken together, our experiments highlight complementary roles played by fluid flows and network strength properties of substrates on which biofilms growth. For growing *Serratia marcescens* colonies, increased substrate stiffness enhances biofilm growth rates. While biofilms may be applying additional traction stresses through adhesion and friction, the experimentally observed substrate deformations are consistent with poroelastic substrate rearrangements in areas out ahead of the growing biofilm front (Fig. 6a).

**Discussions and Conclusions**

Bacteria can sense and respond to the mechanical features of their underlying substrate, but the precise mechanisms remain largely unclear. In this article, we investigated the effects of substrate material properties on the biofilm expansion of *Serratia marcescens.* Using polyacrylamide hydrogels of varying composition, we found that substrate stiffness and porosity tune the transport and spread of growing biofilm colonies. Our results indicated that increasing substrate stiffness enhances biofilm expansion rates in the limit of purely elastic substrates, unlike conventional agar growth substrates.

The observed increase in biofilm growth with substrate stiffness might be surprising given that biofilm expansion rates decrease on substrates of increasing agar concentration. However, this result may point to additional biologically relevant parameters, such as the role of dissipative stresses and non-linear substrate stiffness present in agar (Fig. 1), in mediating biofilm growth.

An emerging number of studies indicate that bacteria sense surfaces by translating mechanical cues presented by the surrounding environment into biochemical signals through mechanosensitive signaling pathways (*27-29*) . At the scale of an individual bacteria, there have been several molecular machines identify as potential candidates to read-out mechanical signals, such as the bacteria flagella (*30-32*), pili (*27, 29*), and cell envelope ion-channels (*33-35*). This process allows bacteria to modulate gene expression, cellular differentiation, and virulence factors (*27*) in response to physical changes in their environment.

Our results provide compelling evidence that biofilms can sense and respond to the mechanical properties of their environment beyond single cells and at the collective level. This has important implications for how bacteria alter gene expression profiles and modify biofilm functions in response to mechanical stresses. Our results suggest new models of biofilm growth that explicitly account for the effects of substrate stiffness and poroelastic substrate remodeling.

Finally, our work highlights the need for biofilm experiments on conceptually simple materials to guide our understanding of bacteria's sense of touch.

**Methods**

*Cell culture:* There were four strains of bacteria used in this study*: Serratia marcescens* (274 ATCC), *Pseudomonas aeruginosa*, *Proteus Mirabilis* BB2000, and *Myxococcus xanthus*. Bacteria cells were inoculated and grown in LB medium (except for the case of *M. xanthus* is which CTTYE was used as cell culture medium) with shaking at 37 °C overnight. Cells in the cultures were subsequently mixed, OD600 was measured, and the cultures were back diluted to an OD600 of 0.6. An aliquot of 5 µL of this inoculum was spotted on growth substrates (agar or PAA gels of varying stiffness). Cultures were then maintained at 37 °C for 24 hours.

*Gel preperation:* To prepare hydrogels of varying stiffness, polyacrylamide gels were prepared as described previously (*14, 36*). Briefly, polyacrylamide gels were prepared by mixing together acrylamide, bis-acrylamide, and distilled water at various ratios. Polymerization was initiated by the addition of 1 µL electrophoresis grade tetramethylethylenediamine (TEMED) followed by 10 µL of 2% ammonium per-sulfate (APS). 200 µL of the solution were then pipetted between two glass coverslips, one treated with glutaraldehyde (bottom) and the other with SurfaSil-treated (top) and allowed to polymerize for 30 minutes. Then, the top cover slip was removed from the gels, and the final dimensions of the hydrogel formed a disc, 20 mm in diameter and 1mm in height.

*Gel characterization:* Rheology measurements were performed on a Malvern Panalytical Kinexus Ultra+ rheometer equipped with a 20 mm diameter plate. The elastic gel solutions are polymerized at room temperatures between the rheometer plates at a gap height of 1 mm (30 minutes). The shear modulus was then measured as a function of shear strain from 2-50% at a frequency of 1 rad/sec.

*Substrate preparation and inoculation:* To prepare PAA substrates for inoculation, we followed a protocol previously described by Tuson et al (*17*). The PAA gels were washed three times (two ten-minute washes and one overnight wash) in phosphate-buffered saline. The washes were then repeated with LB medium. Before inoculation, the substrates were removed from LB medium, allowed to dry for 20 minutes at room temperature, and then treated with UV sterilization for an additional 20 minutes. The prepared bacterial solution was inoculated onto the center of each gel in a 5 µL droplet. After placing the droplets, 2 µL of liquid was removed from each droplet with a pipette to bring bacteria in closer contact with the gel surface.

*Imaging:* Time-lapse imaging was performed with a Nikon Ti-E inverted microscope equipped with a 4x objective. The cultures were maintained at 37 °C using a Tokai-Hit stage top incubator. Images were taken every 10 minutes for 15 hours using a motorized stage to capture growth at four positions along the edge of each biofilm. After the 15 hours had elapsed, full colony images were taken with a MotiCam camera.

*Motility measurements:* Time lapse images were loaded in custom Python scripts that allowed manual supervision of automated boundary detection (SI materials). The boundaries were fit to circular arcs, and the average length of multiple radial lines connecting subsequent arcs determined the biofilm velocity. Velocities are measured over 20 minute time increments.

*Statistical methods:* Data are presented as mean values ± standard errors (SE). Experiments were conducted in 3+ independent trials. The unpaired Student's *t*-test with two tails at the 95% confidence interval was used to determine statistical significance. Denotations: *, $p < = 0.05$; **, $p < 0.01$; ***, $p < 0.001$; ns, $p > 0.05$.

**Acknowledgements:** We thank Paul Janmey and Roy Welch for insightful discussions. This work was supported by NSF DEB 2033942 and NSF MCB 206747 awards to AP.

**Author contributions:** MA and AP designed the experimental studies. MA performed and analyzed all experimental work. MH contributed to traction force microscopy analysis. AF and RW assisted with *M. xanthus* experiments. AG contributed to gel permeability characterization and poroelastic analysis. MA and AP wrote the manuscript.

**References**

1. C. Prigent-Combaret, O. Vidal, C. Dorel, P. Lejeune, Abiotic surface sensing and biofilm-dependent regulation of gene expression in Escherichia coli. *Journal of bacteriology* **181**, 5993-6002 (1999).
2. Y. Nakamura *et al.*, Establishment of a multi-species biofilm model and metatranscriptomic analysis of biofilm and planktonic cell communities. *Applied microbiology and biotechnology* **100**, 7263-7279 (2016).
3. C. Fei *et al.*, Nonuniform growth and surface friction determine bacterial biofilm morphology on soft substrates. *Proceedings of the National Academy of Sciences* **117**, 7622-7632 (2020).
4. M.-C. Duvernoy *et al.*, Asymmetric adhesion of rod-shaped bacteria controls microcolony morphogenesis. *Nature Communications* **9**, 1120 (2018).
5. F. Beroz *et al.*, Verticalization of bacterial biofilms. *Nature Physics* **14**, 954-960 (2018).
6. J. Yan *et al.*, Mechanical instability and interfacial energy drive biofilm morphogenesis. *eLife* **8**, e43920 (2019).
7. J. Yan, C. D. Nadell, H. A. Stone, N. S. Wingreen, B. L. Bassler, Extracellular-matrix-mediated osmotic pressure drives Vibrio cholerae biofilm expansion and cheater exclusion. *Nature Communications* **8**, 327 (2017).
8. S. Srinivasan, C. N. Kaplan, L. Mahadevan, A multiphase theory for spreading microbial swarms and films. *Elife* **8**, e42697 (2019).
9. L. Ping, Y. Wu, B. G. Hosu, J. X. Tang, H. C. Berg, Osmotic pressure in a bacterial swarm. *Biophysical journal* **107**, 871-878 (2014).
10. A. Seminara *et al.*, Osmotic spreading of <em>Bacillus subtilis</em> biofilms driven by an extracellular matrix. *Proceedings of the National Academy of Sciences* **109**, 1116-1121 (2012).

11. B. Sabass, M. D. Koch, G. Liu, H. A. Stone, J. W. Shaevitz, Force generation by groups of migrating bacteria. *Proceedings of the National Academy of Sciences* **114**, 7266-7271 (2017).
12. A. Cont, T. Rossy, Z. Al-Mayyah, A. Persat, Biofilms deform soft surfaces and disrupt epithelia. *eLife* **9**, e56533 (2020).
13. A. Doostmohammadi, S. P. Thampi, J. M. Yeomans, Defect-Mediated Morphologies in Growing Cell Colonies. *Physical Review Letters* **117**, 048102 (2016).
14. T. Yeung *et al.*, Effects of substrate stiffness on cell morphology, cytoskeletal structure, and adhesion. *Cell Motil Cytoskeleton* **60**, 24-34 (2005).
15. M. Dembo, Y. L. Wang, Stresses at the cell-to-substrate interface during locomotion of fibroblasts. *Biophys J* **76**, 2307-2316 (1999).
16. J. P. Butler, I. M. Tolić-Nørrelykke, B. Fabry, J. J. Fredberg, Traction fields, moments, and strain energy that cells exert on their surroundings. *Am J Physiol Cell Physiol* **282**, C595-605 (2002).
17. H. H. Tuson, L. D. Renner, D. B. Weibel, Polyacrylamide hydrogels as substrates for studying bacteria. *Chemical Communications* **48**, 1595-1597 (2012).
18. M. Bertasa *et al.*, Agar gel strength: A correlation study between chemical composition and rheological properties. *European Polymer Journal* **123**, 109442 (2020).
19. K. Arakawa, Rheological properties of hydrogels of agar-agar. *Bulletin of the Chemical Society of Japan* **34**, 1233-1235 (1961).
20. C. E. Kandow, P. C. Georges, P. A. Janmey, K. A. Beningo, Polyacrylamide hydrogels for cell mechanics: steps toward optimization and alternative uses. *Methods in cell biology* **83**, 29-46 (2007).
21. E. E. Charrier, K. Pogoda, R. G. Wells, P. A. Janmey, Control of cell morphology and differentiation by substrates with independently tunable elasticity and viscous dissipation. *Nature Communications* **9**, 449 (2018).
22. C. A. Grattoni, H. H. Al-Sharji, C. Yang, A. H. Muggeridge, R. W. Zimmerman, Rheology and Permeability of Crosslinked Polyacrylamide Gel. *Journal of Colloid and Interface Science* **240**, 601-607 (2001).
23. Y. Abidine *et al.*, Physical properties of polyacrylamide gels probed by AFM and rheology. *EPL (Europhysics Letters)* **109**, 38003 (2015).
24. K. Little, J. Austerman, J. Zheng, K. A. Gibbs, Cell Shape and Population Migration Are Distinct Steps of <em>Proteus mirabilis</em> Swarming That Are Decoupled on High-Percentage Agar. *Journal of Bacteriology* **201**, e00726-00718 (2019).
25. J. Yan, M. D. Bradley, J. Friedman, R. D. Welch, Phenotypic profiling of ABC transporter coding genes in Myxococcus xanthus. *Frontiers in Microbiology* **5**, 1-12 (2014).
26. K. K. Chelvam, L. C. Chai, K. L. Thong, Variations in motility and biofilm formation of Salmonella enterica serovar Typhi. *Gut pathogens* **6**, 1-10 (2014).
27. A. Persat, Y. F. Inclan, J. N. Engel, H. A. Stone, Z. Gitai, Type IV pili mechanochemically regulate virulence factors in Pseudomonas aeruginosa. *Proceedings of the National Academy of Sciences* **112**, 7563-7568 (2015).
28. F. Song, H. Wang, K. Sauer, D. Ren, Cyclic-di-GMP and oprF are involved in the response of Pseudomonas aeruginosa to substrate material stiffness during attachment on polydimethylsiloxane (PDMS). *Frontiers in microbiology* **9**, 110 (2018).

29. Y. F. Inclan *et al.*, A scaffold protein connects type IV pili with the Chp chemosensory system to mediate activation of virulence signaling in Pseudomonas aeruginosa. *Molecular microbiology* **101**, 590-605 (2016).
30. A. E. Patteson, A. Gopinath, M. Goulian, P. E. Arratia, Running and tumbling with E. coli in polymeric solutions. *Scientific Reports* **5**, (2015).
31. P. P. Lele, B. G. Hosu, H. C. Berg, Dynamics of mechanosensing in the bacterial flagellar motor. *Proceedings of the National Academy of Sciences* **110**, 11839-11844 (2013).
32. N. Wadhwa, R. Phillips, H. C. Berg, Torque-dependent remodeling of the bacterial flagellar motor. *Proceedings of the National Academy of Sciences* **116**, 11764-11769 (2019).
33. A. K. Harapanahalli, J. A. Younes, E. Allan, H. C. van der Mei, H. J. Busscher, Chemical Signals and Mechanosensing in Bacterial Responses to Their Environment. *PLoS Pathog* **11**, e1005057 (2015).
34. J. H. Naismith, I. R. Booth, Bacterial mechanosensitive channels--MscS: evolution's solution to creating sensitivity in function. *Annu Rev Biophys* **41**, 157-177 (2012).
35. E. Perozo, A. Kloda, D. M. Cortes, B. Martinac, Physical principles underlying the transduction of bilayer deformation forces during mechanosensitive channel gating. *Nat Struct Biol* **9**, 696-703 (2002).
36. A. E. Patteson *et al.*, Loss of Vimentin Enhances Cell Motility through Small Confining Spaces. *Small* **113**, 1903180-1903110 (2019).